\documentstyle[epsfig,aps,prc]{revtex}
\begin{document}
\title{Many-body perturbation calculation of spherical nuclei with a separable monopole interaction: I. Finite Nuclei}
\author{ P.~Stevenson$^{(a),(b),(c),(d)}$,
  M.~R.~Strayer$^{(b),(c),(d)}$, and J.~Rikovska Stone$^{(a),(e)}$}   
\address{ 
  {\small \em $^{(a)}$ Clarendon Laboratory, Department of Physics, University of Oxford, Oxford, OX1 3PU, United Kingdom} \\
  {\small \em $^{(b)}$ Physics Division, Oak Ridge National
    Laboratory, P.O. Box 2008 Oak Ridge, Tennessee 37831}\\
  {\small \em $^{(c)}$ Department of Physics and Astronomy, University of Tennessee,
    Knoxville, Tennessee 37996} \\
  {\small \em $^{(d)}$ Joint Institute for Heavy Ion Research, Oak
    Ridge National Laboratory, P.O. Box 2008, Oak Ridge, Tennessee 37831} \\
  {\small \em $^{(e)}$ Department of Chemistry and Biochemistry,
    University of Maryland, College Park, Maryland 20742}
}
\date{\today}
\maketitle
\draft
\begin{abstract}
  We present calculations of ground state properties of spherical,
  doubly closed-shell nuclei from $^{16}$O to $^{208}$Pb employing the
  techniques of many-body perturbation theory using a separable density
  dependent monopole interaction.  The model gives
  results in Hartree-Fock order which are of similar quality to other
  effective density-dependent interactions. In addition, second and
  third order perturbation corrections to the binding energy are
  calculated and are found to contribute small, but non-negligible
  corrections beyond the mean-field result.  The perturbation series
  converges quickly, suggesting that this method may be used to
  calculate fully correlated wavefunctions with only second or third
  order perturbation theory. We discuss the quality of the results and
  suggest possible methods of improvement.
\end{abstract}


\section{Introduction}
The central problem of nuclear structure theory is the solution of the
many-body Schr{\"o}dinger equation (MBSE). For Hamiltionans of interest in
the nuclear case, an analytic solution is impossible, and one is
compelled to use some approximation, either in the numerical solution
of the equation or the specification of the Hamiltonian, or both.

Approaching the problem with the aim of using as realistic a
representation of the potential as possible usually means fitting a
combination of a 
meson exchange and phenomenological interaction to low-energy nucleon-nucleon
scattering data and properties of few-body systems. To get good
agreement with experiment both two- and three-body 
forces seem to be necessary.  Recent examples of such potentials
include the Bonn \cite{Mac87}, the Argonne two-body \cite{Wir95} with
Urbanna 3-body\cite{Car83}, Nijmegen \cite{Sto94} and
Moscow \cite{Kuk99} potentials, the last of which also incorporates
quark degrees of freedom. These forces share the property of having a
hard repulsive core which is a natural consequence of meson-exchange.  It is
this hard core which presents the difficulty in solving the MBSE.  For
instance, Hartree-Fock (HF) mean-field calculations with such interactions
result in unbound nuclei.  Treating corrections beyond the
HF approximation order-by-order in perturbation theory is also
unsuccessful since the interactions used are non-perturbative.  One has
to solve the full MBSE numerically in as exact a way as possible using
techniques such as Variational Monte-Carlo\cite{Wir91}, Green's Function Monte
Carlo\cite{Pud97}, the coupled-cluster method\cite{Coe58,Mih98},
and the Fermion Hyper-netted chain model\cite{Fab98}. Using effective 
interactions derived from realistic potentials, no-core shell-model
calculations have been made in light nuclei\cite{Zhe95} and heavier
nuclei close to closed shells have been treated\cite{Cor98}.

The computational difficulty of performing numerically exact solutions
of the MBSE has
limited the techniques to light nuclei, for instance $A=8$ results
have been published recently using the Argonne $v18$ and Urbanna IX
potentials in the GFMC framework\cite{Wir00}.  In this work, it is
seen that although the lightest nuclei are reproduced very well, the
quantitative comparison of theory to data gets worse as $A$
increases. This may be due to the necessarily phenomenological nature
of the three--body potential, a problem which may be overcome with
re-fitting.  On the other hand, it is not obvious that higher-body
forces will not prove necessary or that the concept of a bare
interaction between nucleons is valid for small distances.

Attempts were made in the late sixties primarily by the Kerman group
at MIT to parameterize the NN interaction in such a way that it is weak
in the sense of being perturbative.  Such a weak interaction
allows one to perform Hartree-Fock calculations to obtain a reasonable
approximation to the full wavefunction and then to calculate
corrections in perturbation theory.  While this technique seems very
attractive, the result obtained were only moderately successful at
reproducing experimental data\cite{Ker69,Bre69,Rou69,Zip70,Rii70}, a
fact which was presumed to be due to inadequacies in the potentials used.
The efficacy of developing a suitable interaction when similar, though
more complicated, techniques were available for realistic interactions
has been questioned\cite{Bet71} and no better interaction was
developed. Separable parameterizations, particularly the
quadrupole-quadrupole interaction\cite{Bes61,Bar65}, have retained
currency, as residual interactions \cite{Dev96}.
Even when the interactions are too strong for regular perturbation
theory, separable interactions requiring solution of Br\"uckner
Hartree-Fock equations have proved fruitful\cite{Kwo97} because of
their simplicity.

On the other hand, interactions have been developed which are not
intended for use in the full MBSE, but rather to give good results
with a Hartree-Fock calculation alone.  Good quantitative success came
with the zero-range density-dependent force of Ehlers and
Moszkowski\cite{Ehl72} and Skyrme's interaction\cite{Sky56}, used in
HF calculations by Vautherin and Brink \cite{Vau72} and subsequently
by many others, and also Gogny's finite-range interaction\cite{Dec80}. 
Skyrme's interaction has been particularly successful, in part due to
its simple form,  that
of a delta function, which leads to easy calculation, even of the
exchange part of the force.  This computational simplicity has allowed
extensive study of the properties of nuclei to be made with the Skyrme
interaction across the entire range of nuclei in the periodic table
\cite{Que78,Dob84,Rei95}.  Related somewhat to
the Skyrme-Hartree-Fock model is the Relativistic Mean Field (RMF)
approach\cite{Ser86,Rei89}, which also gives single-particle motion in
a mean field, but as a solution to the Dirac equation as opposed to
the Schr\"odinger equation. The RMF approach has some nice features
such as the natural occurrence of the spin-orbit splitting without
recourse to an assumed spin-orbit interaction.

These mean-field models are inherently single-particle in nature. The
forces used are not intended to be used in the MBSE, nor are
explicit corrections beyond the mean-field part of the framework,
although Skyrme's interaction can be considered as a phenomenological
G-matrix equivalent\cite{Neg72}, and in that sense includes a subset
of possible correlation effects in the mean-field.  Although
appropriate for use in mean-field calculations, Skyrme's interaction
would actually diverge in perturbation theory because of the zero range.
This compels one to use a different interaction to obtain correlation
beyond the mean field than was used to create it. Extra residual
forces are used, such as pairing\cite{pairing}, or shell model
interactions\cite{Bro98b} to allow for 
more general wavefunctions and obtain more accurate reproduction of
the physics, or certain approximations are used such as
RPA\cite{Boh53}, which can describe certain observables,
particularly those of giant resonance states and other forms of
collective motion, but not others. It is thought that correlations
should be particularly important in nuclei at the limits of stability,
where the nucleons nearest the Fermi level couple strongly to the
continuum.

We revisit the idea that it is possible to
parameterize a nuclear interaction in such a way that it is weak
enough with which to perform perturbation theory, thereby allowing
correlated wavefunctions and observables to be calculated across
the entire range of nuclei.  Using the separable ansatz of previous
``weak'' interactions we have developed a density dependent
interaction which we hope will provide some insight into the
correlation structure of nuclear wavefunctions while retaining the
quantitative power of contemporary effective interactions used in
Hartree-Fock. In contrast to previous work, the interaction is
designed to be an effective interaction with parameters fitted to the
properties of finite nuclei within the calculation framework for which
it is intended.

The paper is organized as follows. In Section 2 we briefly review
basics of many-body perturbation theory. The separable interaction,
used in the present work is given and discussed in Section 3. Results
of the calculation for doubly magic nuclei are summarized in Section
4. Derivation of the HF energy and potential is outlined in Appendix
A.

\section{Single particle models and many-body perturbation theory}
For standard perturbation theory to be successful, the Hamiltonian
must be separated into two parts, one which is solvable ($H_0$), and another which
is ``small'' ($H_1$). The Hamiltonian for a many-fermion problem in
which the particles interact via one- and two-body interactions may be
written schematically as
\begin{equation}
  H = \sum_{ij}\langle i|U|j\rangle\,a^\dagger_ia^{}_j + 
  \frac{1}{4}\sum_{ijkl}\langle ij|V|kl\rangle
  a^\dagger_ia^\dagger_ja^{}_la^{}_k.\label{eq:mbham}
\end{equation}
Typically, the one-body part of the Hamiltonian is just the kinetic
energy. This is a unsuitable choice for $H_0$ in the case of nuclei
since the eigenstates of this operator, namely plane wave states, are
not close enough to the exact eigenstates of the
full Hamiltonian and $H_1$ is thus not small.  By adding and
subtracting a one-body term, $U'$, of ones 
choice, the $H_0$ part of the Hamiltonian may be solvable to give a
wavefunction close to the solution of the full problem, thereby making the
corrections from the $H_1$ part small enough for perturbation theory
to succeed. Schematically the Hamiltonian is now split up into two terms;
\begin{equation}
  H = \underbrace{\sum_{ij}\langle i|U+U'|j\rangle\,a^\dagger_ia^{}_j}_{H_0} + 
  \underbrace{\frac{1}{4}\sum_{ijkl}\langle ij|V|kl\rangle
    a^\dagger_ia^\dagger_ja^{}_la^{}_k -\sum_{ij}\langle i|U'|j\rangle\,a^\dagger_ia^{}_j}_{H_1}.
\end{equation}
One practical choice of $U'$ is a simple, analytically-solvable
external potential, typically -- in the nuclear case --  of
harmonic oscillator or Woods-Saxon form. In
such a case, all of the physics of the nuclear interaction is in the
residual part, $H_1$.  Unfortunately, for 
sensible choices of two-body interaction, the residual part of the
Hamiltonian is non-perturbative in the basis obtained from the
solution of the one-body part of the Hamiltonian, and
one has to perform more exact and exacting interaction shell-model
calculations to arrive at a 
meaningful result. Traditionally, this has involved the
diagonalisation of large matrices, or more recently, auxiliary field
Monte-Carlo calculations in the SMMC\cite{Koo97}.

Another popular choice for $U'$ is that of the Hartree-Fock
mean-field.  The HF mean-field potential is usually derived from the full
Hamiltonian by a variational principle.  Viewed in this way, it is the
one-body potential whose occupied eigenstates form the lowest energy
Slater-Determinant many-body wavefunction possible for the full
many-body Hamiltonian.  Using such a one-body potential, some physics
of the two-body 
interaction is included in the single-particle problem. With a
judicious choice of interaction, one ought to be able to produce very
good results, since the approximation of the nucleus as a system of
non-interacting particles in a mean-field is known to be a good one.
Using the HF potential for $U'$ has the added attraction that it 
leads to vanishing first order corrections to the energy in
perturbation theory.  In this
case, using Wick's theorem\cite{Wic50}, one can re-write
(\ref{eq:mbham}) as 
\begin{equation}
  H=\underbrace{E_0 +
    \sum_{i<\epsilon_F}\varepsilon_i:\!a^\dagger_ia_i\!:}_{H_0} + \underbrace{
    \frac{1}{4} \sum_{ijkl}\langle ij|V|kl\rangle:\!a^\dagger_i
    a^\dagger_j a_l a_k\!:}_{H_1}
\end{equation}
where $H_0$ is the HF mean-field Hamiltonian and $H_1$ is the
perturbing Hamiltonian. Note that $H_1$ is just the full two-body
interaction matrix elements with a time-ordered product of creation
and annihilation operators.  The perturbation series for the energy is
ordered by the number of matrix elements of the potential. Usually a
diagrammatic  representation is used\cite{Hug57}. One can write down
the number of diagrams for any particular order of perturbation
theory.  We have evaluated these for the vacuum amplitude up to order
seven. The results are presented in Table \ref{tab:numdiags}.  One
sees that, for the method of direct evaluation of diagrams by order to
be effective, the series must be sufficiently converged by fourth, or
perhaps fifth, order.

\begin{figure}[ht]
\centerline{
\epsfig{file=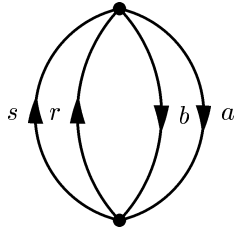}}
  \caption{Labelled second-order Hugenholz diagram\label{fig:seclab}}
\end{figure}

\begin{figure}[ht]
\centerline{
\epsfig{file=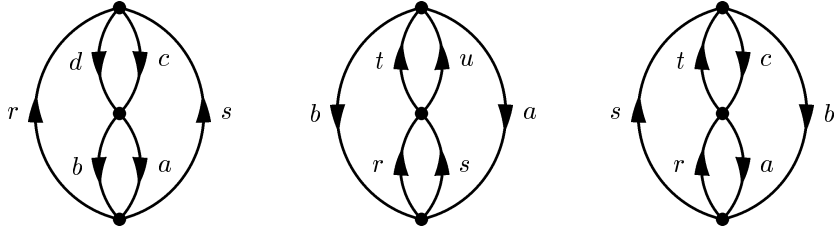}}
  \caption{Labelled third-order Hugenholz diagrams\label{fig:thilab}}
\end{figure}

 In this work, the
second and third 
order diagrams for the vacuum amplitude, as shown in Figs.\
\ref{fig:seclab} and \ref{fig:thilab}, are calculated.  Their algebraic
form is given here as 

\begin{eqnarray}
E_2 &=& \frac{1}{4}\sum\limits_{ab<\epsilon_F} \sum\limits_{rs>\epsilon_F}
    \frac{|\langle ab|\tilde{V}|rs\rangle|^2} 
    {\epsilon_a+\epsilon_b-\epsilon_r-\epsilon_s} \label{eq:2}\\
E_3^{(pp)} &=& \frac{1}{8}\sum\limits_{ab<\epsilon_F}
  \sum\limits_{rs>\epsilon_F} 
  \sum\limits_{tu>\epsilon_F}
    \frac{\langle ab|\tilde{V}|rs\rangle 
      \langle rs|\tilde{V}|tu\rangle 
      \langle tu|\tilde{V}|ab\rangle} 
    {(\epsilon_a+\epsilon_b-\epsilon_r-\epsilon_s) 
      (\epsilon_a+\epsilon_b-\epsilon_t-\epsilon_u)} \label{eq:3pp}\\
E_3^{(hh)} &=& \frac{1}{8}\sum\limits_{ab<\epsilon_F}
  \sum\limits_{cd<\epsilon_F} 
  \sum\limits_{rs>\epsilon_F}
    \frac{\langle ab|\tilde{V}|rs\rangle 
      \langle cd|\tilde{V}|ab\rangle 
      \langle rs|\tilde{V}|cd\rangle} 
    {(\epsilon_a+\epsilon_b-\epsilon_r-\epsilon_s)
      (\epsilon_c+\epsilon_d-\epsilon_t-\epsilon_u)} \label{eq:3hh} \\
E_3^{(ph)} &=& \sum\limits_{abc<\epsilon_F}
  \sum\limits_{rst>\epsilon_F} 
    \frac{\langle ab|\tilde{V}|rs\rangle 
      \langle cr|\tilde{V}|at\rangle 
      \langle st|\tilde{V}|cb\rangle} 
    {(\epsilon_a+\epsilon_b-\epsilon_r-\epsilon_s)
      (\epsilon_b+\epsilon_c-\epsilon_s-\epsilon_t)} \label{eq:3ph}
\end{eqnarray}
in which the tildes over the potential indicate the matrix element is
antisymmetrized.  The state vectors label HF single-particle states,
whose energies are given by the subscripted $\epsilon$.

It is important to note that our interaction is not intended to fit
scattering data, having, as it does, density dependence.  On the level
of the perturbation theory it is necessary to treat the density
functions as just the spatial form of the interaction, rather than a
representation of a many-body force.  This is to be considered a part
of the present model. To do otherwise would be to surrender the
simplifications our weak, separable potential affords.

In the present work, the Hartree-Fock problem is solved in a basis of
spherical harmonic oscillator states.  This yields, along with the
hole states, a large number of particle states, with which to directly
evaluate the sums of the perturbation series.  A sufficient number of
states is used so that the particle states are oscillatory over the
size of the nucleus and that both the HF solution and the perturbation
corrections are reasonably converged. 

\section{Interaction}
\label{sec:interaction}
We have developed an interaction written in the form of a sum
of separable terms, which is to say it is in the form
$V(r_1,r_2)\sim \sum g(r_1)g(r_2)$.  The functions $g$
carry no angular momentum ($l=0$), and the force is dubbed a
monopole-monopole interaction.  For future
applications, it is in intended to include higher multipole
forces, with $l=1,2,\ldots$, within our framework as these will
presumably be necessary for calculation of excited states and deformed
nuclei.  Although higher multipole 
forces will contribute to spherical nuclei from the exchange term in
Hartree Fock order and via correlations in perturbation theory, they
are not included in the present calculation since it seems unwise to attempt
to fit the parameters of such forces to spherical nuclei alone. 

In coordinate space, the monopole interaction is written as
\begin{eqnarray}
  V(\vec{r}_1,\vec{r}_2) &=& W_a f_a \rho^{\beta_a}(\vec{r}_1)\rho^{\beta_a}(\vec{r}_2) (1 + a_a(\tau_1^+\tau_2^-+\tau_1^-\tau_2^+)+b_at_{1z}t_{2z}) \nonumber \\
  &+& W_r f_r \rho^{\beta_r}(\vec{r}_1)\rho^{\beta_r}(\vec{r}_2) (1 + a_r(\tau_1^+\tau_2^-+\tau_1^-\tau_2^+)+b_rt_{1z}t_{2z}) \nonumber \\
  &+& k\nabla_{\!\!1}^2\rho(\vec{r}_1)\nabla_{\!\!2}^2\rho(\vec{r}_2), \label{eq:hamil}
\end{eqnarray}
where the function $f_\xi$ is defined as
\begin{equation}
  f_\xi = \left[\int{\rm d}^3\vec{r}\,
    \rho^{\alpha_\xi}(\vec{r})\right]^{-1}, \label{3:eq:fdef} 
\end{equation}
for subscripts $\xi=a$ and $\xi=r$.  Throughout this work, the three terms in
(\ref{eq:hamil}) are referred to, in the order they appear in the
above expression, as the attractive, repulsive and derivative terms.

In addition, the spin-orbit force is taken to be 
\begin{equation}
  V_{\rm s-o}(r) = c\frac{1}{r}\frac{\partial\rho}{\partial r}\vec{l}
  \cdot \vec{s},
\end{equation}
which is similar to that used in the modified delta interaction \cite{Ehl72}.

The parameters $W_a$, $\alpha_a$, $\beta_a$, $a_a$, $b_a$, $W_r$, $\alpha_r$,
$\beta_r$, $a_r$, $b_r$, $k$ and $c$ are to be fitted to
experimental data.

One notices that the two-body interaction consists of a sum of terms,
each of which is separable in form and that the expressions for the
attractive and repulsive terms in  (\ref{eq:hamil}) differ only by the
values of their parameters

The energy, $E_{\rm pot}$, due to the interaction (\ref{eq:hamil}) in the
Hartree-Fock approximation is derived in Appendix \ref{chap:hfenergy}
(\ref{eq:eatt}), (\ref{eq:nonisoen}), (\ref{eq:ederv}) and is
presented here; 
\begin{eqnarray}
  E_{HF} &=& T + E_{\rm coul} + E_{\rm pot} = T + E_{\rm coul}+\sum_{\xi=a,r}\Big\{\frac{1}{2}W_\xi f_\xi
  N_\xi^2 -\frac{1}{2}W_\xi f_\xi M_\xi +\frac{1}{2}W_\xi b_\xi
  f_\xi(\Delta N_\xi)^2 \nonumber \\
  &&-\frac{1}{2} W_\xi f_\xi [b_\xi M_\xi  + a_\xi
  M_\xi^{(\tau\bar{\tau})}]\Big\}+\frac{1}{2}kN_d^2  + cN_w \label{eq:hfenergy}
\end{eqnarray}
where $T$ is the kinetic energy, $E_{\rm coul}$ is the direct Coulomb
energy plus exchange in the Slater approximation. The following
quantities have been defined; 
\begin{eqnarray}
  N_\xi &=& \int {\rm d}^3\vec{r}\,\rho^{\beta_\xi+1}(\vec{r})\nonumber \\
  M_\xi &=& \int\!\int {\rm d}^3\vec{r}_1{\rm d}^3
  \vec{r}_2 \,\left[\rho_p(\vec{r}_1,\vec{r}_2)\rho^{\beta_\xi}(\vec{r}_1)
    \rho^{\beta_\xi}(\vec{r}_2) \rho_p(\vec{r}_1,\vec{r}_2) +
    \rho_n(\vec{r}_1,\vec{r}_2)\rho^{\beta_\xi}(\vec{r}_1) 
    \rho^{\beta_\xi}(\vec{r}_2) \rho_n(\vec{r}_1,\vec{r}_2)\right]
  \nonumber \\
  \Delta N_\xi &=& \int {\rm d}^3\vec{r}\,\rho^{\beta_\xi}(\vec{r})\left[
    \rho_p(\vec{r})-\rho_n(\vec{r})\right] \nonumber \\
  M_\xi^{(\tau\bar{\tau})} &=& \int\!\int {\rm d}^3\vec{r}_1{\rm d}^3
  \vec{r}_2 \,\left[\rho_p(\vec{r}_1,\vec{r}_2)\rho^{\beta_\xi}(\vec{r}_1)
    \rho^{\beta_\xi}(\vec{r}_2) \rho_n(\vec{r}_1,\vec{r}_2) +
    \rho_n(\vec{r}_1,\vec{r}_2)\rho^{\beta_\xi}(\vec{r}_1) 
    \rho^{\beta_\xi}(\vec{r}_2) \rho_p(\vec{r}_1,\vec{r}_2)\right]
  \nonumber \\
  N_d &=& \int {\rm d}^3\vec{r}\,\rho(\vec{r})\nabla^2\rho(\vec{r})
  \nonumber \\
  N_w &=& \int {\rm d}^3\vec{r}\,\frac{1}{r}\frac{\partial \rho}
  {\partial r} \rho_w(\vec{r})
\end{eqnarray}
and the following densities are used:
\begin{eqnarray}
  \label{eq:den}
  \rho(\vec{r})&=&\rho_p(\vec{r})+\rho_n(\vec{r}) = \sum_{i <\epsilon_F\in
    p}\varphi_i^*(\vec{r})\varphi_i(\vec{r}) + \sum_{i<\epsilon_F\in n}
  \varphi_i^*(\vec{r})\varphi_i(\vec{r}) \\
  \label{eq:denmat}
  \rho(\vec{r}_1,\vec{r}_2) &=& \rho_p(\vec{r}_1,\vec{r}_2) +
  \rho_n(\vec{r}_1,\vec{r}_2) = \sum_{i<\epsilon_F \in p}
  \varphi_i^*(\vec{r}_1)   \varphi_i(\vec{r}_2) +
  \sum_{i<\epsilon_F\in n}  \varphi_i^*(\vec{r}_1)  
  \varphi_i(\vec{r}_2) \\
  \rho_w(\vec{r}) &=& \sum_{i<\epsilon_F} \frac{1}{2}\left( j_i(j_i+1)
    -l_i(l_i+1) - 3/4\right)\varphi^*_i(\vec{r})\varphi_i(\vec{r}). 
\end{eqnarray}
The variation of the total energy is carried out in Appendix
\ref{chap:hfenergy} (see \ref{eq:alphavarterm}, \ref{eq:betavarterm},
\ref{termpropg}, \ref{eq:dervpot}) . The resulting local Hartree-Fock
potential is  
\begin{eqnarray}
  U_{L,\tau}(x) &=& \sum_{\xi=a,r} \Big\{W_\xi f_\xi \left[ N_\xi
    (\beta_\xi+1) + b_\xi\Delta_\tau N_\xi\right]\rho^{\beta_\xi}(x)
  \nonumber \\
  &-& W_\xi(\alpha_\xi/2)f_\xi^2\left[N_\xi^2+b_\xi(\Delta
    N_\xi)^2-(1+b_\xi)M_\xi-a_\xi
    M_\xi^{(\tau\bar{\tau})}\right]\rho^{\alpha_\xi-1}(x)\nonumber \\ 
  &-&W_\xi f_\xi\beta_\xi\left[(1+b_\xi)G_\xi(x)+a_\xi G_\xi^{(\tau
      \bar{\tau})}(x)\right] \rho^{\beta_x-1}(x) \nonumber \\
  &+&\left[W_\xi b_\xi\beta_\xi f_\xi\Delta N_\xi\right]
  \rho^{\beta_x-1}(x)\delta\rho(x)\Big\} \nonumber \\
  &+&2kN_d\nabla^2\rho(x)\label{eq:localpot}
\end{eqnarray}
which differs for protons ($\tau=p$) and neutrons ($\tau=n$) through
the function 
\begin{equation}
  \Delta_\tau N_\xi = \left\{\begin{array}{cl} \Delta N_\xi, & \tau=p\\
      -\Delta N_\xi, & \tau=n.\end{array}\right.
\end{equation}
The other newly-introduced functions in (\ref{eq:localpot}) are
\begin{eqnarray}
  G_\xi(x) &=& G_\xi^{(pp)}(x) + G_\xi^{(nn)}(x) = \int \! {\rm d}^3
  \vec{r}
  \left[\rho_p(\vec{r},\vec{x})\rho^{\beta_\xi}(r)\rho_p(\vec{x},\vec{r})
    + \rho_n(\vec{r},\vec{x})\rho^{\beta_\xi}(\vec{r})\rho_n(\vec{x},\vec{r})\right]
  \nonumber \\
  G^{(\tau\bar{\tau})}_\xi(x) &=& G_\xi^{(pn)}(x) + G_\xi^{(np)}(x) = 
  \int \! {\rm d}^3
  \vec{r}
  \left[\rho_p(\vec{r},\vec{x})\rho^{\beta_\xi}(r)\rho_n(\vec{x},\vec{r})
    + \rho_n(\vec{r},\vec{x})\rho^{\beta_\xi}(\vec{r})\rho_p(\vec{x},\vec{r})\right].\label{eq:geez}
\end{eqnarray}
In addition, the non-local component to the mean field is (see \ref{eq:nonlocterm})
\begin{equation}
  U_{NL,\tau}(\vec{x},\vec{x}') = 
  \sum_{\xi=a,r}W_xf_x
  \rho^{\beta_\xi}(\vec{x})\rho^{\beta_\xi}(\vec{x}')
  \left\{(1+b_\xi)\rho_\tau(\vec{x},\vec{x}') 
    +a_\xi\rho_{\bar{\tau}}(\vec{x},\vec{x}')\right\}.
  \label{eq:nonlocpot}
\end{equation}
and there is a state-dependent potential from the spin-orbit
interaction of the form
\begin{equation}
  U_{\rm so}(\vec{x})\varphi_b(x) = c\left(w_b\frac{1}{x} \frac{\partial
      \rho}{\partial x} - \frac{1}{x}\frac{\partial\rho_w}{\partial x} -
    \frac{1}{x^2}\rho_w(x)\right)\varphi_b(x)
  \label{eq:spinopot}
\end{equation}
where $w_b = 1/2(j_b(j_b+1)-l_b(l_b+1)-3/4)$ is the spin-orbit weight
factor.

Note that the one-body spin-orbit term could be taken as either a
one-body force, or as a one-body potential deriving from a two-body
force. Since the latter approach would render the perturbation
calculation problematic due to the absence of a suitable of form of the
two-body force, we choose the former approach.  Hence, since the
force is density-dependent, we have also included the rearrangement
contribution to the HF 
potential.  Only the non-rearrangement term actually
gives rise to the spin-orbit splittings, but the rearrangement terms,
coming as they do from a variational principle, result in a lowering
of the HF energy. 
Combining the potentials (\ref{eq:localpot}), (\ref{eq:nonlocpot}) and
(\ref{eq:spinopot}) gives us the HF equation
\begin{equation}
  U_{L,\tau}(\vec{x})\varphi_b(\vec{x}) + \int {\rm d}^3\vec{x}'\,\,
  U_{NL,\tau}(\vec{x},\vec{x}')  \varphi_b(\vec{x}') +  U_{\rm
    so}(\vec{x})\varphi_b(\vec{x}) =   \varepsilon_b\varphi_b(\vec{x})
\end{equation}
In this potential, as well as in the expression for the total energy
(\ref{eq:hfenergy}), the exchange contribution from the derivative
term is omitted.  While it would, in principle, be desirable to
include this term, the calculational complexity involved in doing so
has forced the omission in the present case.  However, for the main
attractive and repulsive terms, the exchange part is much smaller than
the direct in all nuclei, and the direct derivative term gives a
rather small contribution to the mean-field and the binding energy in
comparison to the other direct terms, so it is not considered an
unwarranted approximation to neglect the effects of this term.  

It might be objected that the form of the interaction is too
unrealistic. For instance, since it is separable it cannot satisfy
Galilean invariance. Furthermore, the unusual form of the isospin
operators gives rise to a different effective interaction for protons
and neutrons.  As for the second point, since this is a
density-dependent interaction, the effect of the difference between
neutrons and protons comes from the densities as well as the
isospin operators.  This being the case, it seems reasonable to allow
for the ``stretching'' of the isospin operator as we have done, to
allow greater freedom in fitting the interaction to data.  As for the
first issue, the unusual form has been motivated by the desire to keep
the force ``weak'' and is justified by the quality of the results
produced.  A possible method for restoring Galilean invariance within
our framework is discussed in Section \ref{sec:exper}.

The choice of omitting a spin-spin ($\sigma\cdot\sigma$) type force
yet having an 
isospin-isospin type force is motivated by the nuclei under study. All
the closed-shell nuclei are spin-saturated and would contribute only
through the exchange term in the HF order. For this separable
interaction, the space-exchange terms are rather small and a spin-spin
force would add little to the results. In addition, even if the
effects in closed-shell nuclei are important, it does not seem
reasonable to fit this term to closed-shell nuclei alone. It remains an open
question whether such a force will prove necessary or useful in
open-shell nuclei. 

It is interesting to compare the leading terms in the HF mean field to
that of other models.  The first line of equation (\ref{eq:localpot})
gives us this as
\begin{equation}
  U(x) \sim \sum_{\xi=a,r} C_\xi (f_\xi N_\xi) \rho^{\beta_\xi}(x)
\end{equation}
where $C_\xi$ is a combination of constants.  The product $f_\xi
N_\xi$ is
\begin{equation}
  f_\xi N_\xi = \frac{\int {\rm d}^3\vec{r}\,\rho^{\beta_\xi+1}
    (\vec{r})}{\int {\rm d}^3 \vec{r}\,\rho^{\alpha_\xi}(\vec{r})}. 
\end{equation}
If $\beta_\xi+1 = \alpha_\xi$ then the product $f_\xi N_\xi$ is
constant and the leading mean field terms go like
\begin{equation}
  U(x) \sim {\mathcal C}_a\rho^{\beta_a}(\vec{x})  +{\mathcal C}_r
  \rho^{\beta_r} (\vec{x})
\end{equation}
which, for the special case $\beta_a=1$, are the same as the terms in the
Skyrme and Gogny mean-field proportional to the parameters $t_0$ and $t_3$,
which give the bulk of the binding energy and saturation properties.
In this work, we do not strictly keep $\beta_\xi+1=\alpha_\xi$, thus
allowing for some A-dependence of the coefficients in the mean-field
potential. It has been found, however, that one can not go too far away from
the equality and still obtain reasonable results.

That one can get similar results in a mean-field calculation from two
very different interactions is reflected in the different constitution
of the resulting perturbative part of the Hamiltonian, $H_1$.

\section{Doubly (Semi-) Magic Nuclei}
\label{sec:exper}
In order to find the best set of parameters for the interaction
(\ref{eq:hamil}), calculations have been made of 14 doubly closed-shell
nuclei across 
the periodic table. They are $^{16}$O, $^{34}$Si, $^{40,48}$Ca,
$^{48,56,68,78}$Ni, $^{90}$Zr, $^{100,114,132}$Sn, $^{146}$Gd and
$^{208}$Pb. The nuclei represent a selection of doubly-closed
(sub-)shell nuclei both close to and far from stability. There is
limited experimental information about $^{48}$Ni \cite{Bla00} and
$^{100}$Sn \cite{Sch95}. $^{78}$Ni has yet to be discovered.

The ability to reproduce the properties of such exotic nuclei will be
important for applications of our technique and discrepancies will
help direct refinements.  

A Hartree-Fock code assuming spherical symmetry and representing
wavefunctions in a basis of spherical harmonic oscillator states was
used to calculate uncorrelated wavefunctions.  Perturbation
corrections to the binding energy were directly 
evaluated using the results of the HF calculation.  The results
presented here were obtained in a basis of 12 expansion coefficients
per single-particle wavefunction and iterated until the HF energy had
converged to within 1keV.  The parameters of the force were fitted to
binding energies to second order and charge radii, charge density
distributions, single-particle
energies and spin-orbit splittings to HF order of the
nuclei listed above where experimental data were available, and are
presented in Table  \ref{tab:params}.  

\begin{figure}
\centerline{
\epsfig{file=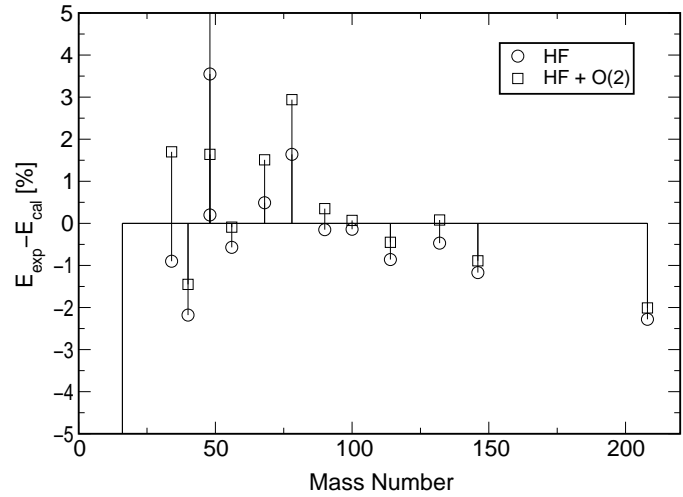,width=8cm,angle=270}}
\caption{ Deviation of Hartree-Fock Energy from experiment. Negative
  errors denote underbinding. Note that $^{16}$O and the second-order result for $^{48}$Ni are beyond the scale.\label{fig:percbind}}
\end{figure}

The results of the the calculated energies, in HF order and in each
order of perturbation theory are presented in Table
\ref{tab:bind}. The differences between the HF energy $E_{\rm cal} =
(\ref{eq:hfenergy}) E_{\rm HF}$ and the experimental ground state
energy $E_{\rm exp}$ and between HF plus second order perturbation
correction (\ref{eq:2}) $E_{\rm cal} = E_{\rm HF}+E_2$ and experiment
are shown in Figure \ref{fig:percbind}.  The experimental energies are
taken from the mass table of Audi and Wapstra \cite{Aud95} with two
exceptions.  An estimate of the mass of the recently-discovered
nuclei $^{48}$Ni \cite{Bro98a} and the measured mass of $^{100}$Sn \cite{Cha96}.
The energy for $^{78}$Ni was taken from \cite{Aud95} in which
extrapolated values are given, which are thought to be in error by
less than $0.2\%$.

 One sees from Figure \ref{fig:percbind} that most of the nuclei fit
 the binding energy to within 
$\sim$2\%.  The most obvious exception is $^{16}$O which is quite
under-bound.  This may be due to the omission of a center-of-mass
correction which would undoubtedly go a long way to close the
discrepancy in the energy\cite{Ben00}.  It was not calculated in this
case since a rigorous microscopic correction would destroy the
mean-field which provides the essential basis for the perturbation
calculation.  A phenomenological correction could have been
calculated, but perhaps the most suitable method in our framework
would be to include with our multipole forces an isoscalar dipole
force which could be fitted to restore the translational invariance of
the many-body Hamiltonian, and evaluated exactly in perturbation theory.

A general trend can be seen in which lighter nuclei are somewhat
over-bound and the heaviest are under-bound.  It is the exceptions which
conspire to stop the fitting algorithm from doing better, but the
somewhat systematic nature of this discrepancy suggests that a better
mass or isospin dependence may improve matters.  It is unclear as yet
the extent to which multipole correlations or a spin-spin force
would improve the fit to spherical nuclei.  That question awaits the
study of deformed nuclei and excited states.  

  In Table
\ref{tab:skcomp} a comparison is made of the quality of the 
fit to the binding energy to properties of the same nuclei calculated
with a selection of Skyrme parameterizations. The parameterizations used are
SIII\cite{Bei75}, SkP\cite{Dob84}, SLy4\cite{Cha95} and
SkI4\cite{Rei95}. 
In this comparison, it is seen that the energies
from the  different Skyrme parameterizations are of a similar quality,
all reproducing the binding energies of closed-shell nuclei very
well, with only a few binding energies being reproduced no better than
1\% -- including $^{48}$Ni whose experimental value is in any case not
well known. It is clear that the results from the separable force are
somewhat worse. Particularly problematic is $^{16}$O, whose large
under-binding was mentioned above, and also $^{48}$Ni which is, as
with the Skyrme parameterizations, over-bound, although more so with
the separable interaction. 

Results for one-body properties are also presented. Comparison of the charge
density results to experiment \cite{Fri95} and to the selection of Skyrme
interactions is made in Table \ref{tab:radii}.
One-body observables are generally reproduced better in the HF
calculation alone than the binding energies.  
The comparison of the calculated charge radii with experiment is
generally more favourable than the energy data. The radius of oxygen
is too large by about 5\% which  is consistent with its under-binding.
It can be seen that the agreement with experiment is of the same level
as the Skyrme interactions. Perturbative corrections to the one-body
observables, such as the densities, and hence radii, will be
calculated in future work.  

\begin{figure}[bp]
\centerline{
\epsfig{file=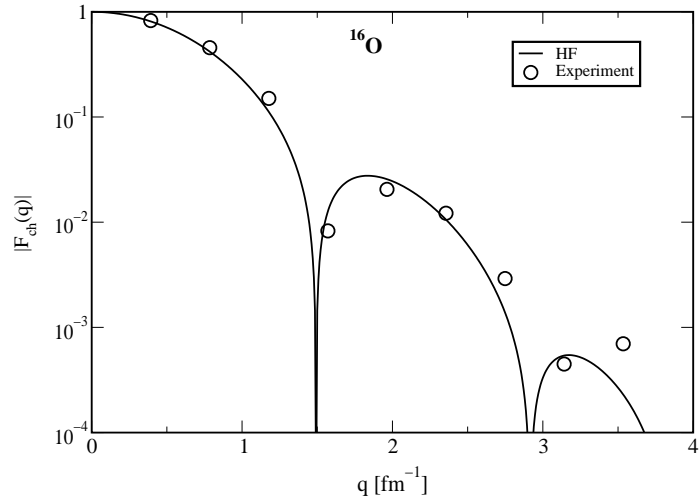,width=8cm,angle=270}}
\caption{  Charge Form factor in $^{16}$O\label{fig:o16ff}}
\end{figure}

\begin{figure}[bp]
\centerline{
\epsfig{file=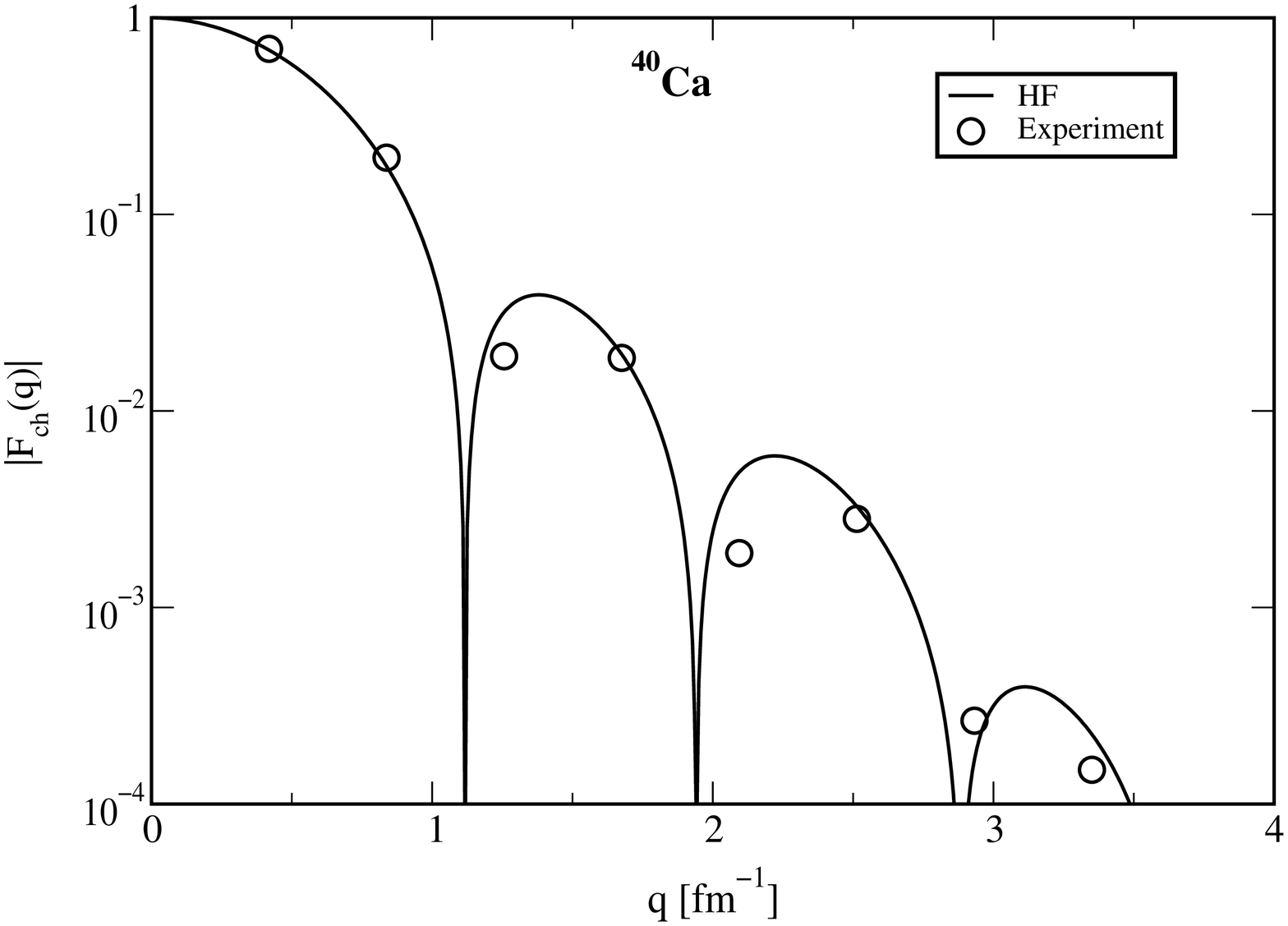,width=8cm,angle=270}}
\caption{  Charge Form factor in $^{40}$Ca\label{fig:ca40ff}}
\end{figure}

\begin{figure}[bp]
\centerline{
\epsfig{file=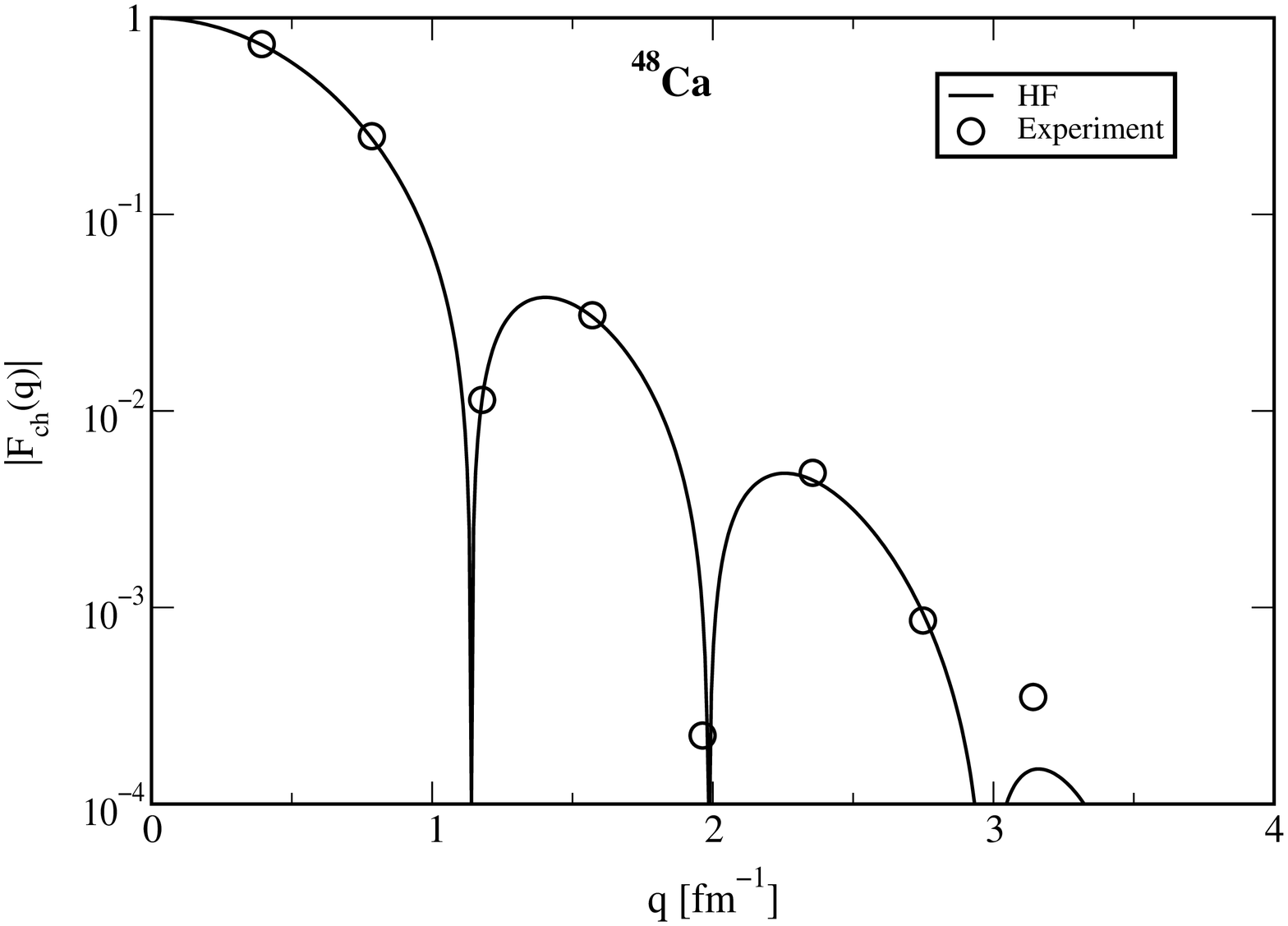,width=8cm,angle=270}}
\caption{  Charge Form factor in $^{48}$Ca\label{fig:ca48ff}}
\end{figure}

\begin{figure}[bp]
\centerline{
\epsfig{file=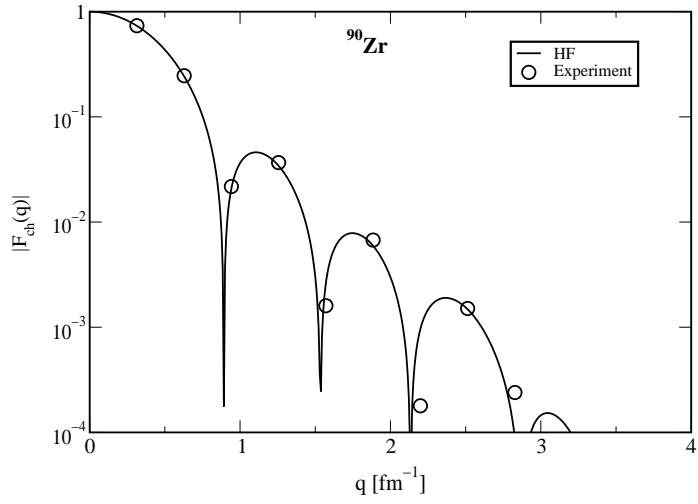,width=8cm,angle=270}}
\caption{  Charge Form factor in $^{90}$Zr\label{fig:zr90ff}}
\end{figure}

\begin{figure}[bp]
\centerline{
\epsfig{file=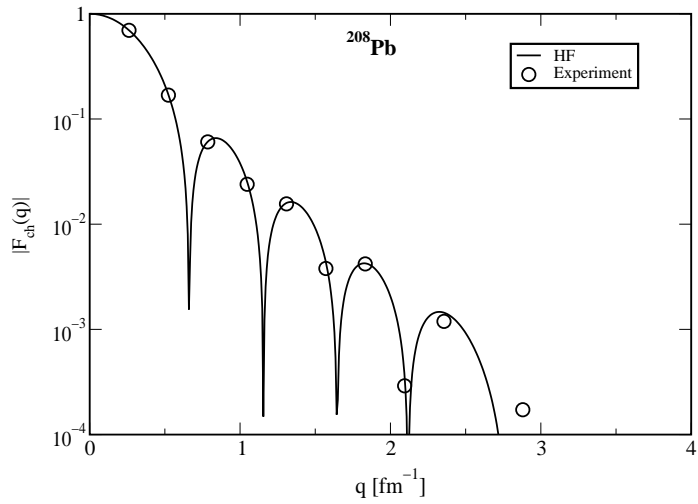,width=8cm,angle=270}}
\caption{  Charge Form factor in $^{208}$Pb\label{fig:pb208ff}}
\end{figure}

Figures \ref{fig:o16ff}-\ref{fig:pb208ff} show the electron scattering
form-factors of a selection of the nuclei compared to
experiment \cite{Fri95}.  The proton density was corrected for the
finite proton 
size by folding with a Gaussian to give the charge density, from which
the radii and form-factors were calculated.  The form-factors agree
with experiment quite well, which is expected given the generally
correct radii.

\begin{figure}[bp]
\centerline{
\epsfig{file=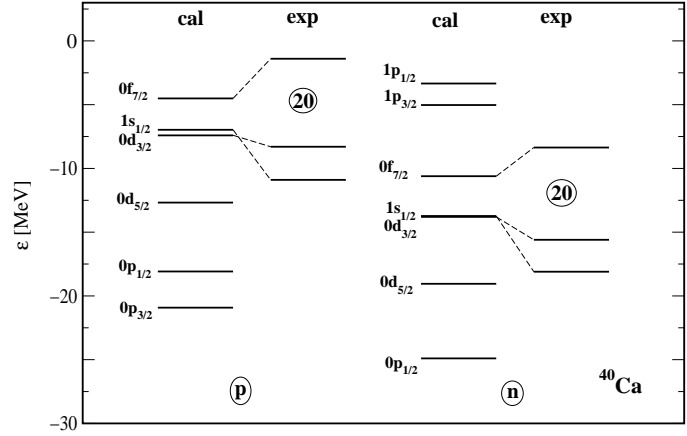,width=8cm,angle=270}}
\caption{  Single-particle energies in $^{40}$Ca compared to
  experiment \protect\cite{Bro98a}\label{fig:ca40spe}}
\end{figure}

\begin{figure}[bp]
\centerline{
\epsfig{file=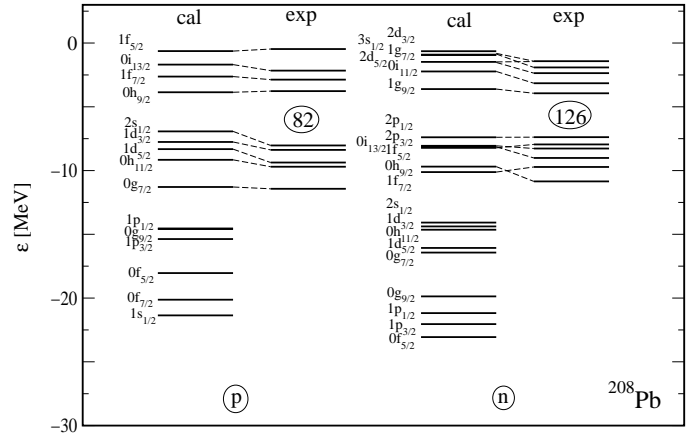,width=8cm,angle=270}}
\caption{  Single-particle energies in $^{208}$Pb compared to
  experiment \protect\cite{Bro98a}\label{fig:pb208spe}}
\end{figure}

Some typical single-particle energies for light and heavy nuclei are
shown in Figures \ref{fig:ca40spe} and \ref{fig:pb208spe} for
$^{40}$Ca and $^{208}$Pb respectively. 
The single-particle energies
of a density-dependent Hartree-Fock calculation do not directly
correspond to an experimental observable, so caution should be used in
comparing values.  It can be seen that the level spacings and shell
closures are better reproduced in $^{208}$Pb, which is true of heavy  nuclei in
general.  The comparatively poorer results in light nuclei seems to be
common to mean-field approaches \cite{Bro98a}.  In the case of
$^{40}$Ca, the gap at the fermi level may be widened by the inclusion
of multipole forces which will link the occupied levels with the
$f_{7/2}$ states, which is otherwise ``inert''.

Table \ref{tab:split} shows spin-orbit splittings for some cases
where the experimental values are known. 
The
``experimental'' data presented represents that 
used in previous work for fitting effective interactions to data
\cite{Cha95,Fri86,Bou87}. Clearly the splittings are all
systematically small.  This could be remedied by an increase in the
spin-orbit coefficient, $c$. In a previous work \cite{Vau68}, a value
10\% higher than ours was used for the same spin-orbit interaction,
and hence the spin-orbit splittings were more realistic.
The lower value used in our work is the result of a compromise between
the reproduction of the spin-orbit splittings and the total binding
energies.   This is a further indication that a more suitable
spin-orbit potential needs to be sought.

The perturbation correction to the energy are seen to be
rather small in all nuclei considered. This is consistent with our goal that
the mean field solution should be close to the exact solution of the
MBSE.  The size of the second order correlation is roughly constant
across the periodic table. It is characterized by a dimensionless strength
parameter, $\kappa$, defined as
\begin{equation}
\kappa = \frac{1}{4}\sum_{ab\le\epsilon_F}
  \sum_{rs>\epsilon_F} \frac{\langle ab|\tilde{V}|rs\rangle 
    \langle rs|\tilde{V}|ab\rangle}{\left(\epsilon_a+\epsilon_b-\epsilon_r
    -\epsilon_s\right)^2}.
\end{equation}
It is related to the ``wound integral''\cite{Bra66} and is
proportional to the number of 2p2h states excited due to the second
order perturbation. 

\begin{figure}
\centerline{
\epsfig{file=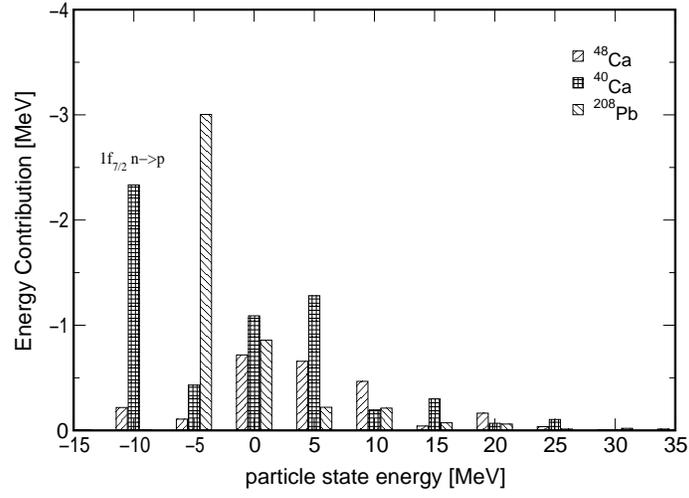,width=8cm,angle=270}}
\caption{ Second-Order Ground-State Correlation structure in $^{40}$Ca,
  $^{48}$Ca and $^{208}$Pb.\label{fig:calcorr}}
\end{figure}

Figure \ref{fig:calcorr} shows the correlation structure from the
second order correction in the $N=Z$ nucleus $^{40}$Ca, and the nuclei
$^{48}$Ca and $^{208}$Pb.  The contribution to the second-order energy
is defined as a function of one of the particle states $r$;
\begin{equation}
  E_2(|r\rangle) = \frac{1}{4}\sum_{ab\le \epsilon_F}
  \sum_{s>\epsilon_F} \frac{\left|\langle ab|\tilde{V}|rs\rangle
  \right|^2 } {\epsilon_a+\epsilon_b-\epsilon_r-\epsilon_s}.
\end{equation}
The plot shows the contribution to the
total second order energy correction as a function of the single
particle energy $\epsilon_r$, in 5 Mev wide bins.  In all three cases
particles are dominantly excited to low-lying states above the Fermi
level. This results in a ground state with occupation probabilities
similar to those which result from pairing forces.  It is also a
further indication that perturbation theory makes sense for our
interaction since it does not predict excitation of particles in the
ground state to extremely high energies.

Since the
calculations were made only using a monopole force, the correlation
structure is not expected to be complete.  Only corrections involving
simultaneous $l=0$ scattering of two particles is included.  An
indication of this is 
seen in the difference between the results for $N=Z$ and $N\ne Z$
nuclei.  The second order correction in $^{48}$Ca is much larger than
that in $^{40}$Ca, due to the possibility of an $f_{7/2}$ neutron
exciting to the $f_{7/2}$ proton state while another proton excites to
a neutron state. This extra excitation is the labelled peak in Fig.\ \ref{fig:calcorr}.  As well as having large wavefunction overlaps, the
energy denominator in this case is much smaller than in any other
possible monopole excitation which must excite any particle 
across major shells to keep all angular and isospin quantum numbers
the same.  When general excitations are permitted by higher
multipole forces $l=1,2\ldots$, this difference between correlations
in $N=Z$ and $N\ne Z$  nuclei will be smoothed out.  For this reason, too, the
correlation energies should not be considered too quantitatively at
this stage, but rather as an indication of the perturbative properties of the
interaction. 

Comparing the form of the interaction to that of Skyrme suggests
other possible sources of improvement to the model. One such  may come
from a  better parameterization of the spin-orbit interaction.  A two-body
form which fits the philosophy of the separable effective interaction
has not been found, but may be necessary to give the correct
contribution to the binding energy. In any case, the simple form used
in the present work which depends on a radial derivative will need
modification if it is to be applied to deformed nuclei. It may also
prove fruitful to 
explore a more general term dependent upon the derivatives of the
density than the single term with parameter $k$, such as is found in
the Skyrme interaction with two terms proportional to $t_1$ and $t_2$,
which often carry further exchange parameters $x_1$ and $x_2$.

\section{Conclusion}
We have presented a new effective nuclear interaction which is
designed for use in calculations which go beyond the mean-field.  The
technique of using perturbation theory to build correlations on top of
the Hartree-Fock result is applicable to our interaction and results
in small corrections to the single-particle behavior.  A
monopole-monopole force alone gives reasonable results for the ground
state properties of spherical doubly-magic nuclei.  It is expected
that the addition of multipole forces will improve these results,
particularly through the completion of the correlation structure.
Such multipole forces will also presumably be important in giving the
correct shapes of deformed nuclei, which are the subject of a
forthcoming study and in the correct description of excited states.

\section{Acknowledgements}
The authors would like to acknowledge useful discussions with D.~Vautherin, P.-G.~Reinhard, D.~M.~Brink and D.~J.~Dean.
This research was sponsored by the Division of Nuclear Physics, U.S. Dept. of 
Energy under contract DE-AC05-00OR 22725 managed by UT--Battelle, LLC, and 
by research the UK EPSRC,  and by US DOE grant no. DE-FG02-94ER40834.
\appendix
\newpage
\section{HF Energy and Potential}
\label{chap:hfenergy}
The interaction is given in equation (\ref{eq:hamil}).  Its
expectation value is the contribution it makes to the total energy and
is
\begin{equation}
  E_{\rm pot} = \frac{1}{2}\sum_{ij<\epsilon_F}\langle
  ij|V\left\{|ij\rangle - |ji\rangle\right\}.
\end{equation}

If we consider just the {\em attractive} term - that is, the term
whose parameters have the subscript ``a'' - then we will obtain the
contribution from the repulsive term by simply substituting the
subscript ``a'' for ``r''.
\begin{eqnarray}
  E_{\rm a} &=& \frac{1}{2}W_af_a\sum_{ij<\epsilon_F}\langle ij |
  \rho^{\beta_a}(r_1)\rho^{\beta_a}(r_2)\left\{|ij\rangle -
    |ji\rangle\right\} \nonumber\\
  &+&\frac{1}{2}W_af_aa_a\sum_{ij<\epsilon_F}\langle ij|
  \rho^{\beta_a}(r_1) \rho^{\beta_a}(r_2) (\tau_1^+\tau_2^- +
  \tau_1^-\tau_2^+) \left\{|ij\rangle -
    |ji\rangle\right\} \nonumber \\
  &+&\frac{1}{2}W_af_ab_a\sum_{ij<\epsilon_F}\langle ij|\rho^{\beta_a}(r_1) \rho^{\beta_a}(r_2) t_{1z}t_{2z}\left\{|ij\rangle -  |ji\rangle\right\}.
  \label{eq:eatt}
\end{eqnarray}
Taking the first line, the matrix element is represented in space (and
spin and isospin) coordinates;
\begin{eqnarray}
  &&\frac{1}{2}W_af_a\sum_{ij<\epsilon_F}\int {\rm d}^3\vec{r}_1
  \int {\rm d}^3\vec{r}_2\,\,
  \phi^*_i(\vec{r}_1) \phi^*_j(\vec{r}_2) \rho^{\beta_a}(\vec{r}_1)
  \rho^{\beta_a}(\vec{r}_2)\phi_i(\vec{r}_1)\phi_j(\vec{r}_2) \nonumber
  \\
  &&-\frac{1}{2}W_af_a\sum_{ij<\epsilon_F}\int {\rm d}^3\vec{r}_1
  \int {\rm d}^3\vec{r}_2\,\,
  \phi^*_i(\vec{r}_1) \phi^*_j(\vec{r}_2) \rho^{\beta_a}(\vec{r}_1)
  \rho^{\beta_a}(\vec{r}_2)\phi_j(\vec{r}_1)\phi_i(\vec{r}_2)\nonumber
  \\
  &=&\frac{1}{2}W_af_a\int{\rm d}^3\vec{r}_1\,\,\rho^{\beta_a+1}(\vec{r}_1)
  \int{\rm d}^3\vec{r}_2\,\,\rho^{\beta_a+1}(\vec{r}_2)\nonumber \\
  &&-\frac{1}{2}W_af_a\int{\rm d}^3\vec{r}_1\int{\rm d}^3\vec{r}_2\,\, 
  \rho_p(\vec{r}_1,\vec{r}_2)\rho^{\beta_a}(\vec{r}_1)\rho^{\beta_a}(\vec{r}_2)
  \rho_p(\vec{r}_2,\vec{r}_1)\nonumber \\
  &&-\frac{1}{2}W_af_a\int{\rm d}^3\vec{r}_1\int{\rm d}^3\vec{r}_2\,\, 
  \rho_n(\vec{r}_1,\vec{r}_2)\rho^{\beta_a}(\vec{r}_1)\rho^{\beta_a}(\vec{r}_2)
  \rho_n(\vec{r}_2,\vec{r}_1)\nonumber \\
  &=&\frac{1}{2}W_af_aN_a^2-\frac{1}{2}W_af_aM_a \label{eq:nonisoen}
\end{eqnarray}
where quantities defined in section \ref{sec:interaction} are
used. Note that integrals include sums over spinors and isospinors
where appropriate, and the coordinates include
spin and isospin coordinates where appropriate.  Where densities are
used, the summing over isospin states has already been done and
where densities do not carry isospin labels, the isoscalar density is
assumed. See equations (\ref{eq:den})
and (\ref{eq:denmat}) for definitions.

The second line of (\ref{eq:eatt}) contains ``isospin-flipping''
operators whose action is to turn an isospin state where particle one
is a proton and particle two a neutron, $|pn\rangle$,  into
$|np\rangle$ and vice-versa. The direct contribution, in which the
labels in the bra and the ket are in the same order is zero since
all proton states are orthogonal to all neutron states.  The exchange
term is similar to that in (\ref{eq:nonisoen}) but with different
isospin combination of the density matrices;
\begin{eqnarray}
  &&-\frac{1}{2}W_af_aa_a\int{\rm d}^3\vec{r}_1\int{\rm d}^3\vec{r}_2\,\, 
  \rho_p(\vec{r}_1,\vec{r}_2)\rho^{\beta_a}(\vec{r}_1)\rho^{\beta_a}(\vec{r}_2)
  \rho_n(\vec{r}_2,\vec{r}_1)\nonumber \\
  &&-\frac{1}{2}W_af_aa_a\int{\rm d}^3\vec{r}_1\int{\rm d}^3\vec{r}_2\,\, 
  \rho_n(\vec{r}_1,\vec{r}_2)\rho^{\beta_a}(\vec{r}_1)\rho^{\beta_a}(\vec{r}_2)
  \rho_p(\vec{r}_2,\vec{r}_1)\nonumber \\
  &=&-\frac{1}{2}W_af_aa_aM_a^{(\tau\bar{\tau})}.
\end{eqnarray}
The third line of \ref{eq:eatt} contains isospin-projection operators
which have a value $+1$ when $i$ and $j$ are like particles and $-1$
when they are unlike.  In the direct term this gives an energy of
\begin{eqnarray}
  &&\frac{1}{2}W_af_ab_a\int{\rm d}^3\vec{r}_1\,\rho_p(\vec{r}_1)
  \rho^{\beta_a}(\vec{r}_1) \int{\rm d}^3\vec{r}_2\,\rho_p(\vec{r}_2)
  \rho^{\beta_a}(\vec{r}_2) \nonumber \\
  &&+ \frac{1}{2}W_af_ab_a\int{\rm d}^3\vec{r}_1\,\rho_n(\vec{r}_1)
  \rho^{\beta_a}(\vec{r}_1) \int{\rm d}^3\vec{r}_2\,\rho_n(\vec{r}_2)
  \rho^{\beta_a}(\vec{r}_2) \nonumber \\
  &&-\frac{1}{2}W_af_ab_a\int{\rm d}^3\vec{r}_1\,\rho_p(\vec{r}_1)
  \rho^{\beta_a}(\vec{r}_1) \int{\rm d}^3\vec{r}_2\,\rho_n(\vec{r}_2)
  \rho^{\beta_a}(\vec{r}_2) \nonumber \\
  &&-\frac{1}{2}W_af_ab_a\int{\rm d}^3\vec{r}_1\,\rho_n(\vec{r}_1)
  \rho^{\beta_a}(\vec{r}_1) \int{\rm d}^3\vec{r}_2\,\rho_p(\vec{r}_2)
  \rho^{\beta_a}(\vec{r}_2) \nonumber \\
  &=&\frac{1}{2}W_af_ab_a\left[\int{\rm d}^3\vec{r}\,\rho_p(\vec{r})
    \rho^{\beta_a}(\vec{r})\right]^2+\frac{1}{2}W_af_ab_a\left[\int{\rm
      d}^3\vec{r}\,\rho_n(\vec{r}) \rho^{\beta_a}(\vec{r})\right]^2
  \nonumber \\
  &&-W_af_ab_a\int{\rm d}^3\vec{r}_1\,\rho_n(\vec{r}_1)
  \rho^{\beta_a}(\vec{r}_1) \int{\rm d}^3\vec{r}_2\,\rho_p(\vec{r}_2)
  \rho^{\beta_a}(\vec{r}_2)\nonumber \\
  &=&\frac{1}{2}W_af_ab_a\left[\int{\rm d}^3\vec{r} (\rho_p(\vec{r}) -
    \rho_n(\vec{r}))\rho^{\beta_a}(\vec{r}) \right]^2 = \frac{1}{2} W_a
  f_a b_a (\Delta N_a)^2.
\end{eqnarray}
The exchange term gives a contribution only when $i$ and $j$ have the
same isospin quantum number, which is just like the case for having no
isospin operator there, so the contribution is like that in
(\ref{eq:nonisoen})
\begin{equation}
  -\frac{1}{2}W_af_ab_aM_a.
\end{equation}

The HF mean-field is obtained by varying the total energy with respect
to the  single particle states. This gives, for the case of the
attractive term, without the explicit isospin dependence:
\begin{eqnarray}
\frac{\delta}{\delta\phi_b(\vec{x})}\left(\frac{1}{2}W_af_aN_a^2 -
  \frac{1}{2}W_af_aM_a\right) &=&\frac{1}{2}W_a\frac{\delta f_a}
  {\delta \phi^*_b(\vec{x})}N_a^2 + W_af_aN_a\frac{\delta N_a} {\delta
  \phi^*_b(\vec{x})} \nonumber \\
&-&\frac{1}{2}W_a\frac{\delta f_a}{\delta \phi^*_b(\vec{x})}M_a -
  \frac{1}{2} W_a f_a \frac{\delta M_a}{\delta \phi^*_b(\vec{x})}.
\label{eq:varnoniso}
\end{eqnarray}
The variation of the function $f_a$ is given by
\begin{eqnarray}
\frac{\delta f_a}{\delta \phi^*_b(\vec{x})} &=& \frac{\delta}{\delta
  \phi^*_b(\vec{x})} \left[\int \rho^{\alpha_a}(\vec{r}){\rm
  d}^3\vec{r}\right]^{-1} \nonumber \\
&=&-f_a^2\int\,{\rm d}^3\vec{r}\,\frac{\rho^{\delta
  \alpha_a}(\vec{r})} {\delta\phi^*_b(\vec{x})} \nonumber \\
&=&-f_a^2\alpha_a\int\,{\rm d}^3\vec{r}\,\rho^{\alpha_a-1}(\vec{r})
  \frac{\delta \rho(\vec{r})}{\delta\phi_b^*(\vec{x})} \nonumber \\
&=& -f_a^2\alpha_a\rho^{\alpha_a-1}(\vec{x})\phi_b(\vec{x})
\end{eqnarray}
so that the contributions of the two terms in (\ref{eq:varnoniso})
involving the variation of $f_a$ give a contribution to the HF
mean--field of 
\begin{equation}
-W_a(\alpha_a/2)f_a^2(N_a^2-M_a)\rho^{\alpha_a-1}(x)
\label{eq:alphavarterm}
\end{equation}
The function $N_a$ is similar in form to $f_a$ and the functional
variation proceeds in a similar manner;
\begin{eqnarray}
\frac{\delta N_a}{\delta \phi_b^*(\vec{x})} &=& \frac{\delta } {\delta
  \phi^*_b(\vec{x})} \int\,{\rm d}^3\vec{r}\,\rho^{\beta_a+1}(\vec{r})
  \nonumber \\
&=& (\beta_a-1)\rho^{\beta_a}(\vec{x})\phi_b(\vec{x})
\end{eqnarray}
and the contribution from the second term in (\ref{eq:varnoniso}) to
the mean-field is
\begin{equation}
W_af_aN_a(\beta_a-1)\rho^{\beta_a}(\vec{x}).
\label{eq:betavarterm}
\end{equation}
Finally, the functional variation of the exchange matrix element $M_a$
is
\begin{eqnarray}
\frac{\delta M_a}{\delta\phi^*_b(\vec{x})} &=& \frac{\delta}{\delta\phi^*_b(\vec{x})}\left(\sum_{ij<\epsilon_F}\int\int {\rm d}^3 \vec{r}_1 {\rm d}^3 \vec{r}_2 \phi^*_i(\vec{r}_1)\phi^*_j(\vec{r}_2)\rho^{\beta_a}(\vec{r}_1)\rho^{\beta_a}(\vec{r}_2)\phi_j(\vec{r}_1)\phi_i(\vec{r}_2)\right)\nonumber \\
&=& 2\sum_{ij<\epsilon_F}\int\int {\rm d}^3 \vec{r}_1 {\rm d}^3 \vec{r}_2 \frac{\delta\phi^*_i}{\delta\phi^*_b(\vec{x})}\phi_j(\vec{r}_2)\rho^{\beta_a}(\vec{r}_1)\rho^{\beta_a}(\vec{r}_2)\phi_j(\vec{r}_1)\phi_i(\vec{r}_2)\nonumber \\
&+& 2\sum_{ij<\epsilon_F}\int\int {\rm d}^3 \vec{r}_1 {\rm d}^3 \vec{r}_2 \phi^*_i(\vec{r}_1)\phi^*_j(\vec{r}_2)\frac{\delta\rho^{\beta_a}(\vec{r}_1)}{\delta\phi^*_b(\vec{x})}\rho^{\beta_a}(\vec{r}_2)\phi_j(\vec{r}_1)\phi_i(\vec{r}_2)\nonumber \\
&=&2\sum_{j<\epsilon_F}\int {\rm d}^3\vec{r}_2\phi^*_j(\vec{r}_2)\rho^{\beta_a}(\vec{x})\rho^{\beta_a}(\vec{r}_2)\phi_j(\vec{x})\phi_b(\vec{r}_2)\nonumber \\
&+&2\sum_{ij<\epsilon_F}\int {\rm d}^3 \vec{r}_2
\phi^*_i(\vec{x})\phi_j(\vec{r}_2)\beta_a\rho^{\beta_a-1}(\vec{x})\phi_b(\vec{x})\rho^{\beta_a}(\vec{r}_2)\phi_j(\vec{x})\phi_i(\vec{r}_2)\label{eq:varm}
\end{eqnarray}
where use is made of the symmetry of the integral to combine the four
terms into two.  The last term in (\ref{eq:varm}) gives rise to a
local term in the mean-field of
\begin{eqnarray}
&&-W_af_a\beta_a\sum_{i<\epsilon_F} \left(\int\,{\rm d}^3\vec{r}\,
  \rho(\vec{r},\vec{x})\rho^{\beta_a}(\vec{r})
  \rho(\vec{x},\vec{r})\right) \rho^{\beta_a-1}(\vec{x})\nonumber \\
&=&-W_af_a\beta_a G_a(\vec{x})\rho^{\beta_a-1}(\vec{x})
\label{termpropg}
\end{eqnarray}
where $G_a(\vec{x})$ has been defined as in (\ref{eq:geez}).

The other term in (\ref{eq:varm}) gives rise to a truly non-local Fock
term in the mean-field:
\begin{equation}
U(\vec{x},\vec{x}')\phi_b(\vec{x}) = -W_af_a\sum_{i<\epsilon_F}
\rho^{\beta_a}(\vec{x})\phi_i(\vec{x}) \left[\int\,{\rm d}^3\vec{r}\,
  \phi^*_i(\vec{r})\rho^{\beta_a}(\vec{r})\phi_b(\vec{r})\right].
\label{eq:nonlocterm}
\end{equation}

This completes the non-isospin-dependent part of the attractive force,
and so also the repulsive by change of subscript.  The
isospin-dependent terms are obtained in an analogous way, except that
when the variation applies to the density of a single nucleon species,
so the contribution to the mean-field applies only to that species.

For the final term in equation (\ref{eq:hamil}), the so-called {\em
  derivative} term, only the direct part of the energy is at present
considered. It is
\begin{eqnarray}
  E_{\rm deriv.} &=& \frac{1}{2}k\sum_{ij<\epsilon_F}\langle
  ij|\nabla^2_{\!1} \rho(\vec{r}_1) \nabla^2_{\!2} (\vec{r}_2) | ij
  \rangle \nonumber \\
  &=&\frac{1}{2}k\int{\rm d}^3\vec{r}_1\,\rho(\vec{r}_1)\nabla^2_{\!1}
  \rho(\vec{r}_1) \int{\rm d}^3\vec{r}_2\,\rho(\vec{r}_2) \nabla^2_{\!2}
  \rho(\vec{r}_2)= \frac{1}{2}kN_d^2. \label{eq:ederv}
\end{eqnarray}
The functional variation proceeds as
\begin{eqnarray} 
\frac{\delta E_{\rm deriv .}}{\delta \phi^*_b(\vec{x})} &=& kN_d
    \frac{\delta}{\delta \phi^*_b(\vec{x})}\int\,{\rm d}^3\vec{r}\,
    \rho(\vec{r}) \nabla^2\!\rho(\vec{r}) \nonumber \\
&=&kN_d\int\,{\rm d}^3\vec{r}\, \left\{
    \frac{\delta \rho(\vec{r})}{\delta\phi^*_b(\vec{x})}\right\}
    \nabla^2\!\rho(\vec{r}) + kN_d\int\,{\rm d}^3\vec{r} \,
    \rho(\vec{r}) \left\{ \frac{\delta}{\delta \phi^*_b(\vec{x})}
    \nabla^2 \! \rho(\vec{r})\right\}.
\end{eqnarray}
The first term gives a contribution to the mean field of 
\begin{equation}
 kN_d\nabla^2\!\rho(\vec{x}).
\end{equation}
By integrating the second term by parts twice, one in fact gets exactly
the same contribution to the mean-field again, so that the total
contribution to the mean-field from the direct term of the derivative
interaction is
\begin{equation}
2kN_d\nabla^2\!\rho(\vec{r})
\label{eq:dervpot}
\end{equation}
The exchange part of this term is not calculated.

\newpage

\begin{table}
  \caption{Number of Hugenholz diagrams by order of perturbation theory\label{tab:numdiags}}
  \begin{tabular}{rlllllll}
    Order & 2 & 3 & 4 & 5 & 6 & 7\\
    \hline
    Labelled Diagrams & 1 & 3 & 39 & 840 & 27,300 & 1,232,280 \\
    Cumulative Diagrams & 1 & 4 & 43 & 883 & 28,183 & 1,260,463 \\
  \end{tabular}
\end{table}

\begin{table}[hb]
\caption{ Monopole force parameters\label{tab:params}}
\begin{tabular}{ccccc}
$W_a$ & $\alpha_a$ & $\beta_a$ & $a_a$ & $b_a$ \\
-1543.8 MeV fm$^3$ & 2.0 & 1.0 & -0.4295 & -0.444825  \\
\hline
$W_r$ & $\alpha_r$ & $\beta_r$ & $a_r$ & $b_r$ \\
1778.0 MeV fm$^{3.8265}$& 2.2165 & 1.246 & -1.4788 & -0.314625 \\ 
\hline
\multicolumn{2}{c}{$c$} && $k$& \\
\multicolumn{2}{c}{160.0 Mev fm$^5$}&& 16.0 Mev fm$^{10}$&\\
\end{tabular}
\end{table}

\begin{table}
  \caption{Hartree-Fock energy (\protect\ref{eq:hfenergy}) and corrections
  from perturbation theory
  (\protect\ref{eq:2})-(\protect\ref{eq:3ph}) compared with 
  experimental value from \protect\cite{Aud95}. All energies are in MeV\label{tab:bind}} 
  \begin{tabular}{rrrrrrrrr}
    \multicolumn{1}{c} {Nucleus} & \multicolumn{1}{c} {$E_{HF}$} &
    \multicolumn{1}{c} {$E^{(2)}$} & \multicolumn{1}{c} {$E^{(3)_{hh}}$} &
    \multicolumn{1}{c} {$E^{(3)_{pp}}$} &
    \multicolumn{1}{c} {$E^{(3)_{ph}}$} & \multicolumn{1}{c}  {$E_{HF+2+3}$} & \multicolumn{1}{c} {$\kappa$} & \multicolumn{1}{c} {Expt.} \\ 
    \tableline
  $^{16}$O  & -109.32&-3.31&-0.1365&-0.3624&+0.921&-112.21 &0.063& -127.68\\
  $^{34}$Si & -280.88&-7.37&-0.0384&-0.4830&+1.223&-287.55 &0.232& -283.43\\
  $^{40}$Ca & -334.53&-2.51&-0.0323&-0.1114&+0.233&-336.95 &0.052& -342.00\\
  $^{48}$Ca & -417.01&-5.97&-0.0189&-0.2725&+0.273&-422.70 &0.202& -416.16\\
  $^{48}$Ni & -360.69&-6.57&-0.0130&-0.2058&+0.427&-367.05 &0.234& -348.33\\
  $^{56}$Ni & -481.25&-2.31&-0.0210&-0.0643&+0.123&-483.52 &0.046& -483.99\\
  $^{68}$Ni & -593.33&-6.00&-0.0109&-0.2091&+0.484&-598.85 &0.221& -590.43\\
  $^{78}$Ni & -651.90&-8.34&-0.0053&-0.1458&+0.477&-659.92 &0.342& -641.38\\
  $^{90}$Zr & -782.70&-3.91&-0.0070&-0.1257&+0.103&-786.51 &0.149& -783.89\\
  $^{100}$Sn& -825.65&-1.71&-0.0060&-0.0220&+0.048&-827.35 &0.039& -826.81\\
  $^{114}$Sn& -963.20&-4.04&-0.0046&-0.1093&+0.226&-967.12 &0.162& -971.57\\
  $^{132}$Sn&-1097.65&-6.17&-0.0023&-0.0864&+0.209&-1103.70&0.287&-1102.92\\
  $^{146}$Gd&-1190.32&-3.42&-0.0026&-0.0699&+0.142&-1193.66&0.146&-1204.44\\
  $^{208}$Pb&-1599.04&-4.51&-0.0013&-0.0664&+0.108&-1603.51&0.233&-1636.45\\
  \end{tabular}
\end{table}

\begin{table}
\caption{Percentage error in binding energy. Negative values are
    underbound.  Separable force is HF+perturbations, Skyrme
    calculations are HF+pairing.\label{tab:skcomp}}
  \begin{tabular}{rrrrrr}
    \multicolumn{1}{c} {Nucleus} & \multicolumn{1}{c} {Sep.} &
    \multicolumn{1}{c} {SIII}    & \multicolumn{1}{c} {SkP}  &
    \multicolumn{1}{c} {SLy4}    & \multicolumn{1}{c} {SkI4} \\
\tableline
  $^{16}$O  & -12.10 &  0.36 & -0.12 &  0.19 &  0.57\\
  $^{34}$Si & 1.45   &  0.43 &  0.92 &  1.08 &  1.06\\
  $^{40}$Ca & -1.49  & -0.18 &  0.21 &  0.50 &  0.51\\
  $^{48}$Ca & 1.61   &  0.40 & -0.04 & -0.63 &  0.31\\
  $^{48}$Ni &  5.37  &  1.52 &  1.22 &  0.74 &  1.58\\
  $^{56}$Ni & -0.10  & -0.22 & -1.11 &  0.09 & -0.29\\
  $^{68}$Ni & 1.43   & -0.25 &  0.09 &  0.81 &  0.26\\
  $^{78}$Ni & 2.89   &  0.58 & -0.05 & -0.29 &  0.21\\
  $^{90}$Zr & 0.33   & -0.14 & -0.14 & -0.11 &  0.13\\
  $^{100}$Sn& 0.07   &  0.14 & -0.51 &  0.75 &  0.28\\
  $^{114}$Sn& -0.46  & -0.71 & -0.48 &  0.09 & -0.56\\
  $^{132}$Sn& 0.07   &  0.10 & -0.31 &  0.14 & -0.12\\
  $^{146}$Gd& -0.90  & -0.33 & -0.41 & -0.20 & -0.24\\
  $^{208}$Pb& -2.01  & -0.17 & -0.27 &  0.21 & -0.24\\
  \end{tabular}
\end{table}

\begin{table}
\caption{Spin-Orbit splittings in HF calculation.  For source of
    experimental values, see text.\label{tab:split}}
  \begin{tabular}{rrr}
    levels & splitting (HF) & splitting (exp) \\
    \tableline
$^{16}$O, 0p (p) & 4.2 & 6.3 \\
$^{16}$O, 0p (n) & 4.3 & 6.1 \\
$^{40}$Ca, 0d (p) & 5.3 & 7.2 \\
$^{40}$Ca, 0d (n) & 5.3 & 6.3 \\
$^{208}$Pb, 2p (n) & 0.67 & 0.89 \\
\end{tabular}
\end{table}
\begin{table}
\caption{Comparison of charge radii between experiment, the separable
  interaction and a selection of Skyrme interactions. The
  model-dependent experimental values are from \protect\cite{Fri95}
  \label{tab:radii} }
  \begin{tabular}{rrrrrrr}
    Nucleus & exp. & Sep. & SIII & SkP  & SLy4 & SkI4 \\
  $^{16}$O  & 2.69 & 2.85 & 2.71 & 2.80 & 2.76 & 2.72 \\ 
  $^{34}$Si & --   & 3.19 & 3.23 & 3.25 & 3.23 & 3.21 \\ 
  $^{40}$Ca & 3.48 & 3.54 & 3.48 & 3.52 & 3.49 & 3.45 \\
  $^{48}$Ca & 3.48 & 3.47 & 3.52 & 3.53 & 3.51 & 3.45 \\
  $^{48}$Ni & --   & 3.88 & 3.77 & 3.82 & 3.79 & 3.80 \\
  $^{56}$Ni & 3.78 & 3.84 & 3.80 & 3.80 & 3.78 & 3.74 \\
  $^{68}$Ni & --   & 3.89 & 3.94 & 3.93 & 3.91 & 3.81 \\
  $^{78}$Ni & --   & 3.87 & 4.02 & 3.99 & 3.98 & 3.97 \\
  $^{90}$Zr & 4.27 & 4.29 & 4.31 & 4.30 & 4.28 & 4.23 \\
  $^{100}$Sn& --   & 4.60 & 4.53 & 4.52 & 4.50 & 4.45 \\
  $^{114}$Sn& 4.60 & 4.65 & 4.66 & 4.62 & 4.62 & 4.59 \\
  $^{132}$Sn& --   & 4.66 & 4.78 & 4.74 & 4.73 & 4.70 \\
  $^{146}$Gd& 4.96 & 5.02 & 5.03 & 5.00 & 4.99 & 4.94 \\
  $^{208}$Pb& 5.50 & 5.50 & 5.57 & 5.52 & 5.51 & 5.48 \\
\end{tabular}
\end{table}

\newpage

\end{document}